\documentclass[aps,twocolumn,pra,superscriptaddress,amsmath,showpacs,tightenlines,pdflatex,longbibliography]{revtex4-1}
\usepackage{amssymb}
\usepackage{amsmath}
\usepackage{dcolumn}
\usepackage{graphicx}
\usepackage{mathrsfs}
\usepackage{appendix}
\usepackage{graphicx}
\usepackage{booktabs}
\usepackage{color}

 \def\note#1{#1}

\usepackage{url}
\usepackage[colorlinks]{hyperref}
\hypersetup{%
    plainpages=true,
    breaklinks=true,
    hypertexnames=false,
    pageanchor=true,
    colorlinks=true,
    linkcolor={blue},
    citecolor={blue},
    urlcolor={blue},
    anchorcolor={black}
}

\begin{document}

\title{Hybrid quantum device with a carbon nanotube and a flux qubit for
dissipative quantum engineering}

\author{Xin Wang}
\affiliation{Institute of Quantum Optics and Quantum Information,
School of Science, Xi'an Jiaotong University, Xi'an 710049, China}
\affiliation{CEMS, RIKEN, Wako-shi, Saitama 351-0198, Japan}

\author{Adam Miranowicz}
\email{miran@amu.edu.pl}
\affiliation{CEMS, RIKEN, Wako-shi, Saitama 351-0198, Japan}
\affiliation{Faculty of Physics, Adam Mickiewicz University,
61-614 Pozna\'n, Poland}

\author{Hong-Rong Li}
\email{hrli@mail.xjtu.edu.cn}
\affiliation{Institute of Quantum Optics and Quantum Information,
School of Science, Xi'an Jiaotong University, Xi'an 710049, China}

\author{Franco Nori}
\email{fnori@riken.jp}
\affiliation{CEMS, RIKEN, Wako-shi, Saitama 351-0198, Japan}
\affiliation{Physics Department, The University of Michigan, Ann
Arbor, Michigan 48109-1040, USA}
\date{\today}

\begin{abstract}
We describe a hybrid quantum system composed of a micrometer-size
carbon nanotube (CNT) longitudinally coupled to a flux qubit. We
demonstrate the usefulness of this device for generating
high-fidelity nonclassical states of the CNT via dissipative
quantum engineering. Sideband cooling of the CNT to its ground
state and generating a squeezed ground state, as a mechanical
analogue of the optical squeezed vacuum, are two additional
examples of the dissipative quantum engineering studied here.
Moreover, we show how to generate a long-lived
macroscopically-distinct superposition (i.e., a Schr\"odinger
cat-like) state. This cat state can be trapped, under some conditions, in a dark state, as
can be verified by detecting the optical response of control fields.
\end{abstract}

 \maketitle

\affiliation{Institute of Quantum Optics and Quantum Information,
School of Science, Xi'an Jiaotong University, Xi'an 710049, China}
\affiliation{CEMS, RIKEN, Wako-shi, Saitama 351-0198, Japan}

\affiliation{CEMS, RIKEN, Wako-shi, Saitama 351-0198, Japan}
\affiliation{Faculty of Physics, Adam Mickiewicz University,
61-614 Pozna\'n, Poland}

\affiliation{Institute of Quantum Optics and Quantum Information,
School of Science, Xi'an Jiaotong University, Xi'an 710049, China}

\affiliation{CEMS, RIKEN, Wako-shi, Saitama 351-0198, Japan}
\affiliation{Physics Department, The University of Michigan, Ann
Arbor, Michigan 48109-1040, USA}

\section{Introduction}

The quantum engineering of nanomechanical systems, which enables
generating nonclassical states of their mechanical motion, has a
variety of applications, such as exploring the classical-quantum
boundary~\cite{Bassi2003,Bassi13}, high-precision
metrology~\cite{Munro02, Giovannetti2011,Volkoff16}, and quantum
information processing~\cite{Stannigel12}. It has been extensively
discussed in hybrid platforms, such as optomechanical and
eletromechanical systems~\cite{Blencowe04,
Meystre2012,rXiang13,Aspelmeyer14}. However, mechanical
oscillators dissipate energy when exposed to a noisy environment
at finite temperature~\cite{Schwab05,Poot12}, and it is still
challenging to generate their nonclassical states with high
purity. To overcome this problem, \emph{dissipative engineering of
long-lived nonclassical states of mechanical motions} has been
extensively studied (see, e.g.,
Refs.~\cite{Porras12,Everitt14,Abdi16} and references therein).
This is the subject of this paper.

Recently, mechanical resonators made of carbon nanotubes
(CNTs)~\cite{Sazonova04,Huttel09,Steele09,Laird11,
Moser14,Benyamini14,Stadler14,Yao00,Javey04,Yu2012} have attracted
considerable attention due to their distinctive advantages. These
include: high-frequency
oscillations~\cite{Huttel09,Steele09,Laird11}, small mass, and a
high-quality factor up to several
millions~\cite{Huttel09,Moser14}. Since CNTs have a large
current-carrying capacity~\cite{Yao00,Javey04,Yu2012}, it is
possible to couple them with superconducting quantum
circuits~\cite{rYou05,rClarke08,rDiCarlo10,
rBuluta11,rYou2011,rLucero12,rGeorgescu14}, to form quantum
electromechanical systems. Recent
experiments~\cite{Delbecq11,Viennot14,Ranjan15} have discussed how
to combine these two systems together, e.g., to couple quantum
dots in CNTs with a superconducting quantum interference device
(SQUID) and resonators. The novel stamping technologies
fabricating these two isolated artificial systems together are
rather mature~\cite{Viennot14}. However, to couple the motion of a
CNT with SQUIDs, one needs to fabricate such a CNT into the SQUID
loop and might require strong external magnetic
fields~\cite{Schneider12}.

Even though the coherence time of a CNT can be much longer than
that of a qubit, it is difficult to manipulate or detect the
quantum coherence of phonons in a CNT for two reasons: First, the
position displacement of a CNT is too tiny to be detected
effectively under current experimental techniques, and the energy
of a phonon is much weaker than that of a
photon~\cite{Blencowe04,Neil08,Aspelmeyer14}. Second, it is
difficult to create the strong coupling between phonons and other
systems (for example, detectors).

In this paper, we propose strongly-coupled hybrid systems, where a
current-carrying CNT interacts with a flux qubit via a
longitudinal coupling. In this novel setup, the CNT is separately
suspended above the qubit loop, rather than being fabricated into
the circuits. The decoherence noise can be minimized by operating
the flux qubit at its optimal point and placing the CNT at a
special symmetric position. Based on this platform, we show that
it is possible to employ this strong longitudinal coupling for
dissipative engineering of the CNT via the rapid decay of the
qubit. Examples include cooling to its ground and squeezed states,
and trapping the mechanical motion into long-lived
macroscopically-distinct superpositions with a high fidelity.
Moreover, without adding other setups, we demonstrate that it
should be possible to check for imperfections in the trapped dark
Schr\"odinger cat states by detecting the optical response of the
control fields. Therefore, it is possible to manipulate and
observe quantum features of phonons based on our proposal.

\section{Model}
\begin{figure}[tbph]
\centering \includegraphics[width=7.8cm]{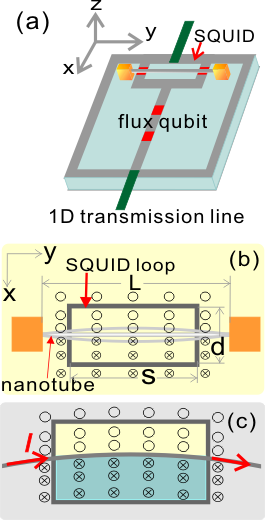}
\caption{Schematic diagrams of the flux qubit hybrid system with a
carbon nanotube (CNT). (a) An eight-shaped flux qubit is placed on
the $x$-$y$ plane with a current-carrying CNT suspended above its
SQUID. The red bars represent the Josephson junctions. \note{The green line is the central
conductor of the 1D transmission line stretched in two directions.} (b)
Specifically: a CNT of length $L$ is located at the central
position along the SQUID (of length $S$ and width
$d$). (c) The dc current $I$, which flows through the CNT \note{(displaced from its equilibrium position)} produces
a flux in the SQUID. The qubit and CNT are coupled
via the motion-induced-flux imbalance between the up (yellow) and
down (blue) sections.} \label{fig1}
\end{figure}

We consider a setup as shown in Fig.~\ref{fig1}(a), in which a
gap-tunable superconducting flux
qubit~\cite{Mooij99,You07,You2008,Fedorov10,Stern14}, is coupled
to a current-carrying CNT~\cite{Sazonova04,Huttel09,Steele09}. The
flux qubit is of an eight-shaped gradiometric topology, which
allows the independent control of the magnetic energy bias
$\epsilon $ and the gap $\Delta $~\cite{Mooij99,You07}. The model
Hamiltonian for the qubit is
\begin{equation}
  H=\tfrac12(\Delta \sigma _{z}+\epsilon \sigma _{x}),
 \label{N1}
\end{equation}
where $\sigma _{z}=|e\rangle\langle e|-|g\rangle\langle g|$, and
$\sigma _{x}=\sigma _{+}+\sigma _{-}=|e\rangle\langle
g|+|g\rangle\langle e|$ are the Pauli operators in the qubit
basis, $|e\rangle$ and $|g\rangle$. For brevity, hereafter, we set
$\hbar=1$. Moreover, $\epsilon =2I_{p}(\Phi _{q}-\Phi _{0}/2)$,
with the flux $\Phi _{q}$ through the qubit and the flux quantum
$\Phi _{0}$, and persistent current $I_{p}$ (see, e.g.,
Refs.~\cite{Fedorov10,Stern14}). The energy gap $\Delta$ is
controlled by the flux $f_{s}$ through the SQUID (of length $S$
and width $d$), and is expressed as
\begin{equation}
 \Delta =\Delta \left( f_{s0}\right) +R\left(\delta\! f_{s}\right),
 \label{N12}
\end{equation}
where $f_{s0}$ is the static flux though the SQUID, $R=\partial
\Delta (f_{s})/(\partial f_{s})$ is the flux sensitivity of the
energy gap~\cite{Paauw09,Zhu2010}, and $\delta\! f_{s}$ describes
the flux perturbations due to external control fields.

As shown in Fig.~\ref{fig1}, the CNT is suspended above the SQUID
in the central-symmetric position and interacts with the SQUID via
the magnetic field produced by its dc current $I$. We assume that
the CNT is longer than the SQUID length, and can be approximately
viewed as a long current line. At its equilibrium position, $x=0$,
the flux contribution of the current $I$ through the SQUID
loop is zero. However, when the CNT starts to vibrate around
$x=0$, the extra flux perturbation due to the area imbalance can
be expressed as $\delta\!f_{s}=\mu _{0}IL\Delta x/(\pi d)$. Thus,
the displacement of the CNT,
\begin{equation}
  \Delta x=\frac{1}{\sqrt{2m\omega _{\text{m}}}}(a+a^{\dag }),
 \label{N2}
\end{equation}
gives rise to a linear modulation of the energy gap of the qubit,
where $m$ ($\omega _{\text{m}}$) is the effective mass (frequency)
of the CNT. Since the CNT is placed symmetrically on the qubit,
the flux contribution from the current $I$ on the energy bias
$\epsilon$ vanishes to first order~\cite{Paauw09,Zhu2010}, i.e.,
the vibration mode decouples from the qubit loop. Moreover, to
minimize the pure dephasing effect, we assume that the flux qubit
is operated at its degeneracy point with $\epsilon =0$ (see, e.g.,
Refs.~\cite{Paauw09,Zhu2010}). We consider driving currents
through a 1D transmission line~\cite{Astafiev2010} [the green line
in Fig.~\ref{fig1}(a)], which produce magnetic fields of opposite
signs in the two-qubit loops~\cite{Schwarz15doc}. Therefore, the
currents interact with the qubit via the dipole matrix element
$\mu=\langle e|MI_{p}\sigma_{x}|g\rangle$, where $M$ is the mutual
inductance. Moreover, the transmission line is placed
symmetrically perpendicular to the CNT, so their mutual inductance
can be zero. Thus, the CNT and the transmission line do not
interact with each other. Assuming that the ac drive current
amplitude $I_{i}$ has frequency $\omega _{i}$, the drive strength
is $2\epsilon _{i}=\mu I_{i}$. The total Hamiltonian becomes
\begin{equation}
H=\frac{\omega _{\text{q}}}{2}\sigma _{z}+ \omega _{\text{m}
}a^{\dag }a+ g\sigma _{z}(a^{\dag }+a)+\sum_{i}
2\epsilon_{i}\sigma_{x}\cos(\omega _{i}t),  \label{eq1}
\end{equation}
where $\omega _{\text{q}} =\Delta \left( f_{s0}\right)$ is the
qubit frequency, and
\begin{equation}
  g= \frac{R\mu_{0}IL}{\pi d\sqrt{2m\omega _{\text{m}}}}
 \label{N9}
\end{equation}
is the qubit-phonon coupling strength. Different from the Rabi
model in standard QED systems, a longitudinal coupling between the
qubit and the CNT is induced~\cite{Wilson10,Liu14,Didier15}. Note
that the Hamiltonian, given in Eq.~(\ref{eq1}), is not at all
specific to the hybrid structure based on the CNT, but only the
numbers discussed in the following paragraphs are specific. For
example, analogous couplings occur in both
circuit-QED~\cite{Didier15} and trapped-ion~\cite{Stenholm86}
systems.  Thus, the study presented here has a much wider
applicability.

We consider that the CNT oscillates at a frequency $\omega
_{\text{m}}/(2\pi )=50~\mathrm{MHz}$, length $5~\mu \mathrm{m}$,
mass $m=4\times 10^{-21}~\mathrm{kg}$ (see, e.g.,
Refs.~\cite{Huttel09, Steele09,Laird11, Moser14,
Benyamini14,LiPB16}), and carries a dc current $I=50~\mu
\mathrm{A}$ (see, e.g., Refs.~\cite{Yao00,Javey04,Yu2012}). For
the qubit, the length and width of the SQUID loop can be about
$3~\mu \mathrm{m}$ and $0.6~\mu \mathrm{m}$, respectively, and the
flux sensitivity of the energy gap~\cite{Paauw09,Fedorov10} can
reach $R=0.7~\mathrm{{GHz}/(m\Phi _{0})}$. Using these parameters,
we obtain the coupling strength $g/(2\pi )=3.4~\mathrm{MHz}$,
i.e., the Lamb-Dicke parameter $\lambda =g/\omega
_{\text{m}}\simeq0.07$. In experiments, the flux sensitivity $R$
and dc current $I$ can be adjusted by changing the gap position
$\Delta \left( f_{s0}\right)$ and the voltage applied to the CNT
gate, respectively. Therefore, the coupling strength $ g $ can be
tuned conveniently.

In realistic situations, we should consider all the decoherence
channels. For a CNT, the quality factor of the vibration mode can
be $\sim\!5\times10^{6}$ (see, e.g., Ref.~\cite{Moser14}). The
dephasing rate of the qubit can be effectively suppressed by
operating at its degeneracy point $\epsilon=0$. Here we employ the
decay channel of the qubit to drive the CNT into nonclassical
states, and the decay rate of a flux qubit around
$\backsim0.5$~MHz is achievable in
experiments~\cite{Bertet05,Stern14}. The mechanical motion of the
CNT might couple to a thermal reservoir with finite temperature
$T$, and the corresponding thermal phonon number is
$n_{\text{th}}=\left[\exp(\omega _{\text{m}
}/k_{B}T)-1\right]^{-1}.$ We assume that the hybrid system is
weakly coupled to a large environment with extremely-short
environmental-memory time, and the Born-Markov approximation is
valid here. The dynamics of this hybrid system can be described by
the Lindblad master equation
\begin{eqnarray}
\frac{d\rho (t)}{dt} &=&-i[H,\rho (t)]+\Gamma D[\sigma _{-}]\rho
(t)  \notag
\\
&&+n_{\text{th}}\gamma D[a^{\dag }]\rho
(t)+(n_{\text{th}}+1)\gamma D[a]\rho (t), \label{eq2}
\end{eqnarray}
where $\rho (t)$ is the time-dependent density matrix of the
hybrid system, and $D[o]\rho =(1/2)(2o\rho o^{\dag }-o^{\dag}o\rho
-\rho o^{\dag }o)$ is the Lindblad superoperator. Assuming that
this hybrid system has a temperature $T\sim 15$~mK, the thermal
phonon number is about $n_{\text{th}}\approx5$. Thus, the coupling
is much stronger than the decoherence of the vibration mode, i.e.,
$g\gg n_{\text{th}}\gamma $.

It should be stressed that the current fluctuation $\delta\!I(t)$
($\ll I$) through the CNT might lead to additional decoherence of
the qubit~\cite{Fedorov10}. Moreover, it might be difficult to
place the CNT at the exact central position of the SQUID. Thus,
the two areas might be slightly imbalanced due to the fabricating
imperfection. Similar with the discussions in
Refs.~\cite{Paauw09,Fedorov10}, the current noise is only
sensitive to the area imbalance, rather than to the whole area of
the loop. In addition, the effects of the flux noise in the SQUID
is much smaller than that for the qubit~\cite{Fedorov10}.
Therefore, the decoherence induced by the current fluctuation
$\delta\!I(t)$ can be effectively suppressed via this spatial
arrangement.
\begin{figure}
\centering \includegraphics[width=8.5cm]{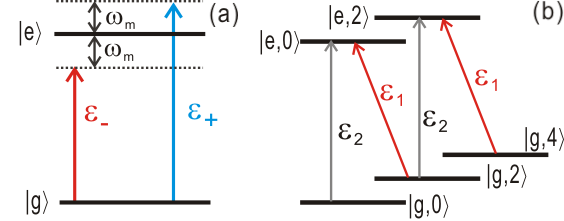} \caption{(Color
online) Schematic diagrams showing how to cool down and engineer
the mechanical mode of the CNT into (a) a squeezed state and (b) a
macroscopically-distinct coherent-state superposition (i.e., a
Schr\"odinger cat-like state).} \label{fig2}
\end{figure}

\section{Cooling the motion to the ground and squeezed ground
states}

Novel ideas about dissipative engineering of a macroscopic motion
into squeezed states, by cooling the Bogoliubov mode into the dark
state, have been proposed in
Refs.~\cite{Rabl04,Tan13,Kronwald13,Wang14}, and recently realized
in microwave optomechanical systems~\cite{Pirkkalainen15}.
Sideband cooling of a mechanical mode into its ground states can
be viewed as a special case of preparing squeezing, by assuming
that the blue-sideband-drive strength is zero~\cite{Marquardt07,
Xue2007, Grajcar08, Schliesser08, Teufel11}. This method creates
stationary squeezed and ground states with the assistance of qubit decay
channels. Here we will show how these methods are employed in this
longitudinal-coupling system. We will show how this method is
employed in this longitudinal-coupling system.

As shown in Fig.~\ref{fig2}, to drive the mechanical mode into the
dark squeezed states, two coherent-drive fields are required,
respectively, of blue and red sidebands with strengths
$\epsilon_{\pm}$  and frequencies $\omega_{\pm}=\omega_{q}\pm
\omega_{\mathrm{m}}$. We obtain the effective Hamiltonian as
\begin{equation}
H_{\text{eff,s}}=\Theta \sigma _{+}B+\text{H.c.}, \label{eq3}
\end{equation}
where $\Theta =2\lambda\sqrt{\epsilon_{-}^{2}-\epsilon_{+}^{2}}$
is the coupling rate and $B$ is the Bogoliubov mode, defined as
\begin{equation}
  B=a^{\dag}\sinh \eta +a \cosh \eta,
 \label{N10}
\end{equation}
with $\tanh \eta=-\epsilon _{+}/\epsilon_{-}$. The slow
decoherence process of the CNT can be neglected, and the effective
Hamiltonian $H_{\text{eff,s}}$, together with the qubit decay
terms in Eq.~(\ref{eq2}) describe the cooling process of the
Bogoliubov mode to its ground state~\cite{Rabl04,Tan13}. It can
easily be verified that the unique stationary state of the system
is $|\Psi _{s}\rangle =|\psi _{s}\rangle |g\rangle$, where
\begin{equation}
  |\psi _{s}\rangle=\exp\left[\tfrac12 (\eta a^{2} -\eta^{*}a^{\dag2})\right]|0\rangle
 \label{N3}
\end{equation}
being the squeezed ground state with squeezing ratio $\eta$. This
is a mechanical analogue of the optical squeezed vacuum.
Therefore, we create stationary squeezed states with the
assistance of decay channels.

By assuming the only nonzero drive to be the red sideband (i.e.,
$\epsilon_{-}>0$ and $\epsilon_{+}=0$), the CNT can be cooled to
its ground state. This effect is similar to the standard sideband
cooling in optomechanical and electromechanical
systems~\cite{Marquardt07,Xue2007,Grajcar08,Schliesser08,Teufel11}.
The steady state for the system is the ground state with no
squeezing, and the effective cooling Hamiltonian reduces to the
standard Jaynes-Cummings model under the rotating-wave
approximation:
\begin{equation}
  H_{\text{eff,s}}=g_{c}\sigma _{+}a+\text{H.c.}
 \label{N4}
\end{equation}
with $g_{c}=2\lambda \epsilon _{-}.$ Assuming $\Gamma \gg g_{c} $,
the excited state of the qubit can be eliminated adiabatically.
The stationary average phonon number satisfies~\cite{Wang14}:
$\bar{n}=(n_{\text{th}}\gamma \Gamma)/(2g_{c}^{2})$.
Since the cooperativity of the system $C=g_{c}^{2}/\gamma
\Gamma\gg1$ is extremely high, the ground state can be achieved
easily with $\bar{n}\sim10^{-3}$ under current experimental
parameters.

\section{Cat-state generation}

Due to rapid decoherence of the mechanical motion, it is still
challenging to observe and coherently manipulate
macroscopically-distinct superpositions (i.e., Schr\"odinger
cat-like states) in realistic
systems~\cite{Bose97,Mancini97,Bose99,Marshall03,RomeroIsart11,
Pepper12,Yin13,Nimmrichter13,Liao16,Abdi16}. To overcome the
decoherence problem, dissipative engineering of a mechanical
resonator into conditional steady superposition states has been
discussed~\cite{Tan132,Asjad14,Everitt14} for quadratic-coupling
optomechanical systems. However, the boson-boson quadratic
coupling is too weak to create observable Schr\"odinger
cat-like states under current experimental
approaches~\cite{Tan132,Asjad14,Thompson08,Sankey10}. Here we show
a novel way to produce a long-lived Schr\"odinger cat-like state
by employing an induced strong quadratic spin-phonon coupling.

As shown in Fig.~\ref{fig2}(b), we applied bichromatic drives for
the qubit: a red-sideband drive with detuning $\approx 2\omega
_{m}$, and a resonant drive with strengths (frequencies) $\epsilon
_{1}$ ($\omega_{1}$) and $\epsilon _{2}$ ($\omega_{2}$),
respectively. Moreover we assume $\epsilon _{2}\ll \epsilon _{1}$.

{By applying the unitary transformation $ U_{1}=\exp \left[
-\lambda \sigma _{z}(a^{\dag }-a)\right] $ to the total
Hamiltonian in Eq.~(\ref{eq1}), we obtain
\begin{equation}
H=\frac{1}{2}\omega _{\text{q}}\sigma _{z}+\omega
_{\text{m}}a^{\dag }a+\sum_{i=1,2}\epsilon _{i}[\sigma
_{+}e^{-i\omega _{i}t}e^{2\lambda (a^{\dag }-a)}+\text{H.c.}].
\label{eqS12}
\end{equation}
Here we assume that $\epsilon _{1}$ is a strong off-resonant drive
strength inducing sideband transitions, and is much stronger than
$\epsilon _{2}.$ To obtain a quadratic coupling, we expand $H$, in
the small parameter $\lambda ,$ to the second and zeroth orders
for the terms $\epsilon _{1}$ and $\epsilon _{2}$, respectively,
and perform the unitary transformation $U=\exp \left( -i \omega
_{1}\sigma _{z}t\right).$ Thus, we obtain
\begin{eqnarray}
H&=&\frac{1}{2}\Delta \sigma _{z}+\omega _{\text{m}}a^{\dag
}a+\epsilon _{i}(\sigma _{+}+\sigma _{-})\nonumber\\ &&+2\epsilon
_{1}[\lambda \sigma _{+}(a^{\dag }-a)+\lambda ^{2}\sigma
_{+}(a^{\dag }-a)^{2}+\text{H.c.}]\nonumber\\  &&+[\epsilon _{2}\sigma
_{+}e^{-i\delta _{12}t}+\text{H.c.}],  \label{eqS13}
\end{eqnarray}%
where $\Delta =\omega _{\text{q}}-\omega _{1}\simeq 2\omega
_{\text{m}}$ is the detuning between the qubit and sideband
drives, and $\delta _{12}=\omega _{2}-\omega _{1}$ is the detuning
between the two drives. The third term induces the dynamical Stark
shift of the qubit. Moreover, due to the coupling and sideband
transition, the frequency of the CNT will also be slightly
renormalized. The shifted frequencies for the qubit and CNT can be
expressed as~\cite{Wang16}:
\begin{equation}
\tilde{\Delta}=\sqrt{\Delta ^{2}+4\Omega _{\text{p}}^{2}},\quad \omega _{m}^{%
{\prime }}=\omega _{m}-\frac{4\epsilon _{1}^{2}g ^{2}}{3\omega
_{0}^{3}}.  \label{eqS14}
\end{equation}
We consider the resonant case $\tilde{\Delta}=2\omega _{m}^{{\prime }%
}=\delta _{12}$. Performing the unitary transformation $U=\exp [-i(\tilde{%
\Delta}\tilde{\sigma}_{z}/2+\omega _{m}^{{\prime }}a^{\dag }a)t]$,
and neglecting all the rapidly-oscillating terms, the Hamiltonian
reduces to
\begin{equation}
H_{\text{eff,c}}=2\epsilon _{1}\lambda ^{2}(\sigma
_{+}a^{2}+\sigma _{-}a^{\dag 2})+\epsilon _{2}(\sigma _{+}+\sigma
_{-}), \label{eqS15}
\end{equation}
where the first term describes the two-phonon sideband
transitions. We can rewrite this effective Hamiltonian as}
\begin{equation}
H_{\text{eff,c}}=(\Theta _{c}\sigma _{+}a^{2}+\epsilon _{2}\sigma
_{+})+ \mathrm{{H.c.},} \label{eq4}
\end{equation}
by denoting $\Theta _{c}=2\lambda^{2}\epsilon _{1}$, which is the
effective two-phonon transition rate. The spin-boson interaction
in Eq.~(\ref{eq4}) is analogous to the purely bosonic coupling in
Refs.~\cite{Tan132,Asjad14}, where the qubit operator is replaced
by those of the cavity field. Since the decoherence of the
high-quality-factor CNT is extremely slow, we only consider the
unitary Hamiltonian in Eq.~(\ref{eq4}) and the qubit rapid-decay
terms in Eq.~(\ref{eq2}), and the dark state for this dissipative
system is $|\Psi _{c}\rangle =|\psi _{c}\rangle |g\rangle,$ where
$|\psi _{c}\rangle$ should satisfy the equation $\left(\Theta
_{c}a^{2}+\epsilon _{2}\right)|\psi _{c}\rangle =0$ (see
Ref.~\cite{Wang17}). If the CNT is initially in an even (odd) Fock
state (e.g., $|0\rangle $ and $|1\rangle$), the steady states are
also even (odd) coherent states (i.e., the famous bosonic
prototypes of Schr\"odinger cat-like states), which are
\begin{equation}
  |\psi _{\alpha ,\pm }\rangle =N^{-1/2}\left( |\alpha \rangle
\pm |-\alpha \rangle \right),
 \label{N5}
\end{equation}
with coherent states $|\pm \alpha \rangle $ ( $\alpha =\sqrt{
-\epsilon _{2}/\Theta _{c}}$), and $|\psi _{\alpha,+}\rangle $
($|\psi _{\alpha,-}\rangle $ ) is the even (odd) coherent states
with $N=2[1+\exp (-2|\alpha |^{2})]$. We consider the cooling
method, which was described in Sec. III, to initially prepare the system into its ground state
$|0\rangle$. After that, the cooling field is shut down and
bichromatic drives are applied. The dark state
$|\psi_{\alpha,+}\rangle$ will be successfully trapped. We define
the fidelity of this target state as $F=\langle \psi _{\alpha
,+}|\rho _{\text{m}}|\psi _{\alpha ,+}\rangle $, where $\rho
_{\text{m}}$ is the reduced density matrix of the mechanical mode.
Moreover, we use the Wigner function,
\begin{equation}
  W(\alpha)=\pi ^{-2}\int d^{2}\beta e^{\alpha \beta ^{\ast
}-\alpha ^{\ast }\beta }\text{Tr}(e^{\beta a^{\dag }-\beta ^{\ast
}a}\rho _{\text{m}}),
 \label{N6}
\end{equation}
to reveal some nonclassical quantum features of the mechanical
states~\cite{Scully1997}. Specifically, we apply the nonclassical
volume of Ref.~\cite{Kenfack04}, which is defined as the
doubled-integrated negative volume of the Wigner function,
\begin{equation}
  \delta_{N}=\int |W(\alpha )|\,d^{2}\alpha-1,
 \label{N7}
\end{equation}
to describe how the superposition interference effects are
different from classical behavior: Higher nonclassical volumes
$\delta_{N}$ indicate more apparent nonclassical features.
\begin{figure}
\centering \includegraphics[width=7.5cm]{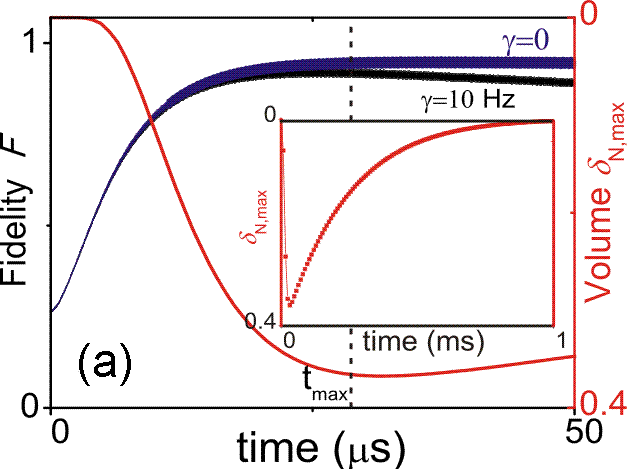}

\includegraphics[width=8cm]{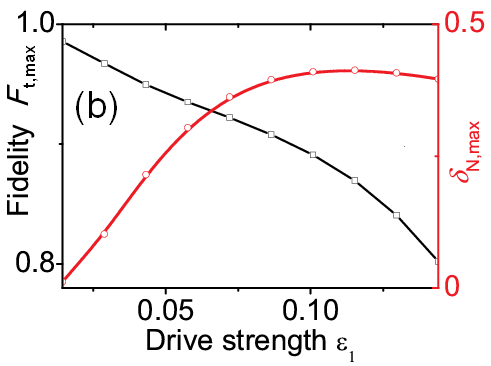}

\caption{(Color online) (a) Time evolutions of the target-state
fidelity $F$ for the mechanical decaying and nondecaying cases,
and the nonclassical volume $\delta_{N}$. The inset shows the
evolution of $\delta_{N}$ in a much longer time scale. (b) The
maximum fidelity $F_{\text{t,max}}$ and the nonclassicality volume
$\delta_{N,\rm{max}}$ versus the resonant driving strength
$\epsilon_{1}$. The vertical dashed line in (a) indicates
$t_{\max}=27.5~\mu\rm{s}$ corresponding to the maximum of $F$. In
panel (a) we assume $\epsilon _{1}/(2\pi )=5~\mathrm{ MHz}$ and
$\lambda =0.06$, corresponding to the quadratic coupling strength
$\Theta _{c}/(2\pi )=2\lambda^{2}\epsilon _{1}=36~\mathrm{KHz}$).
The other parameters are: $\Gamma /(2\pi )=0.4~\mathrm{MHz,}$
$n_{\text{th}}=5$, $\gamma /(2\pi )=10~\mathrm{Hz}$, and $\epsilon
_{2}/(2\pi )=-72~\mathrm{KHz}$.} \label{fig3}
\end{figure}

\begin{figure}
\centering \includegraphics[width=8cm]{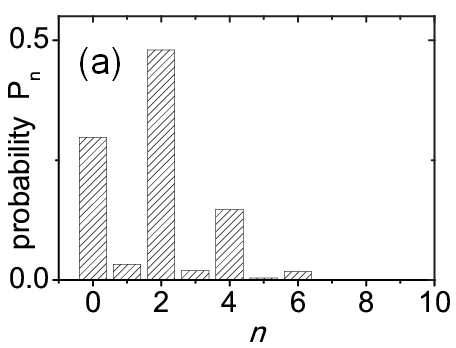}

\includegraphics[width=7.8cm]{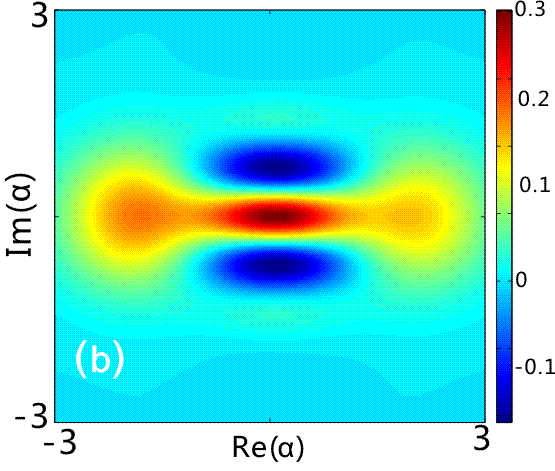}

\caption{(Color online) (a) The phonon-number (Fock-state)
distribution and (b) the Wigner function $W(\alpha)$ for the
Schr\"odinger cat-like state generated at the evolution time
$t_{\max}=27.5~\mu\rm{s}$ [indicated by the vertical dashed line
in Fig.~\ref{fig3}(a)] corresponding to the maximum of the
fidelity $F$. Other parameters are same as in Fig.~\ref{fig3}(a).}
\label{fig4}
\end{figure}

Figure~\ref{fig3}(a) shows the evolution of the fidelity $F$ of
the target superposition states. Without mechanical decay
($\gamma=0$), the fidelity $F$ gradually increases and reaches it
highest value $\sim0.96$, indicating that the mechanical mode is
asymptotically driven into the superposition state $|\psi
_{\sqrt{2},+}\rangle$. The highest fidelity cannot be 1 due to the
high-order terms induced by the off-resonant drive $\epsilon
_{1}$, which are eliminated when deriving the effective
Hamiltonian $H_{\text{eff,c}}$ (see Ref.~\cite{Wang17}). However,
with an extremely slow rate, they would still lead to an even-odd
sideband transition, i.e., $|g\rangle |2n\rangle
\longleftrightarrow |e\rangle |2n\pm1\rangle $. Moreover, by
assuming $n_{\text{th}}=5$ and $\gamma=10~\rm{Hz}$, the thermal
noise will also induce a transition from even to odd Fock states
and vice versa. Compared with the nondecaying case, the fidelity
decreases faster, and $F_{\text{max}}\simeq 0.93$ (at
$t_{\rm{max}}\simeq27~\mu\rm{s}$, the dashed line). However, since
$\gamma$ is extremely slow, the superposition features can last
for a long time.

Moreover, the time evolution of the nonclassical volume
$\delta_{N}$ is plotted in Fig.~\ref{fig3}(a) and in its inset
(with a much longer time scale). The nonclassical volume first
reaches its maximum value $\delta_{N,\rm{max}}$, and then starts
to decrease due to the thermal noise and the deterioration from
the oscillating terms. However, the strength of these processes is
extremely low compared with the preparation rate, the evolution
time $\delta_{N}>0$ is very long ($\sim 1~\rm{ms}$), which might
be enough to detect various nonclassical features of the states.

In Fig.~\ref{fig3}(b), we plotted the highest fidelity
$F_{\text{t,max}}$ and the maximum negative volume
$\delta_{N,\rm{max}}$ versus the resonant drive strength $\epsilon
_{1}$ (i.e., corresponding to the amplitude $\alpha$ of coherent
states $|\pm \alpha \rangle $) with constant quadratic strength
$\Theta _{c}$. As seen from the plot, $\delta_{N,\rm{max}}$
increases with $\epsilon _{1}$ indicating that this nonclassical
signature of the superposition states becomes more apparent. This
is because, with increasing amplitude $\alpha$, the coherent
states $|\alpha \rangle $ and $|-\alpha \rangle $ become nearly
orthogonal, and, thus, more separable and distinct. The
\emph{quantum superposition} features are now more evident.
However, when keeping on increasing $\alpha $, then the
preparation process needs a much longer time, during which thermal
noise can destroy the target states. As a result, the fidelity
$F_{\text{t,max}}$ decreases with increasing $\epsilon _{1}$. When
$\epsilon _{1}/(2\pi )\geq 0.1~\mathrm{MHz} $, both
$F_{\text{t,max}}$ and $\delta_{N,\rm{max}}$ start to decrease due
to these processes. Of course, by reducing the effects of the
thermal environment (by using a CNT with a higher quality factor
or working at lower temperatures), we can choose a larger
$\epsilon _{2}$ to generate more distinct Schr\"odinger cat-like
states.

In Fig.~\ref{fig4}(a), we plotted the phonon-number Fock
distribution for the Schr\"odinger cat-like state generated at
time $t_{\max}$, which is marked by the dashed line in the inset
of Fig.~\ref{fig3}(a). This clearly shows that only the even Fock
states are effectively occupied, while the odd ones have very low
amplitudes. Figure~\ref{fig4}(b) shows the corresponding Wigner
function. We observe two obvious negative regimes and
interference-based evidence for the cat states.

\begin{figure}
\centering
 \hspace{4mm}\includegraphics[width=8.0cm]{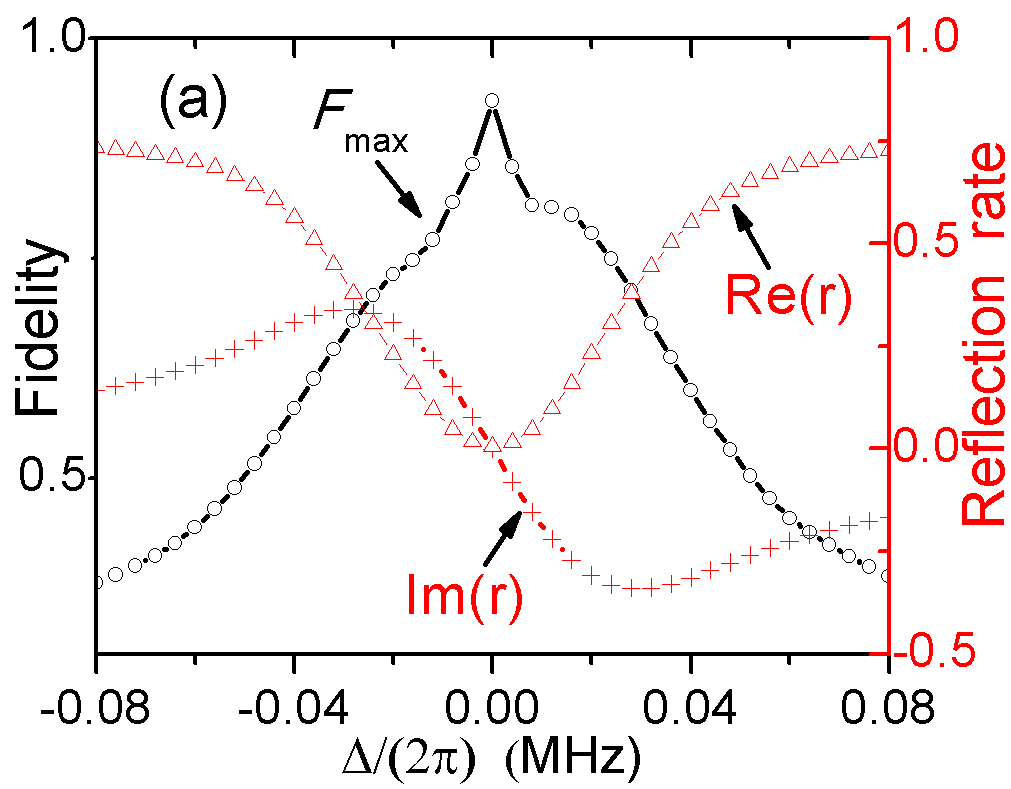}

 \includegraphics[width=8.0cm]{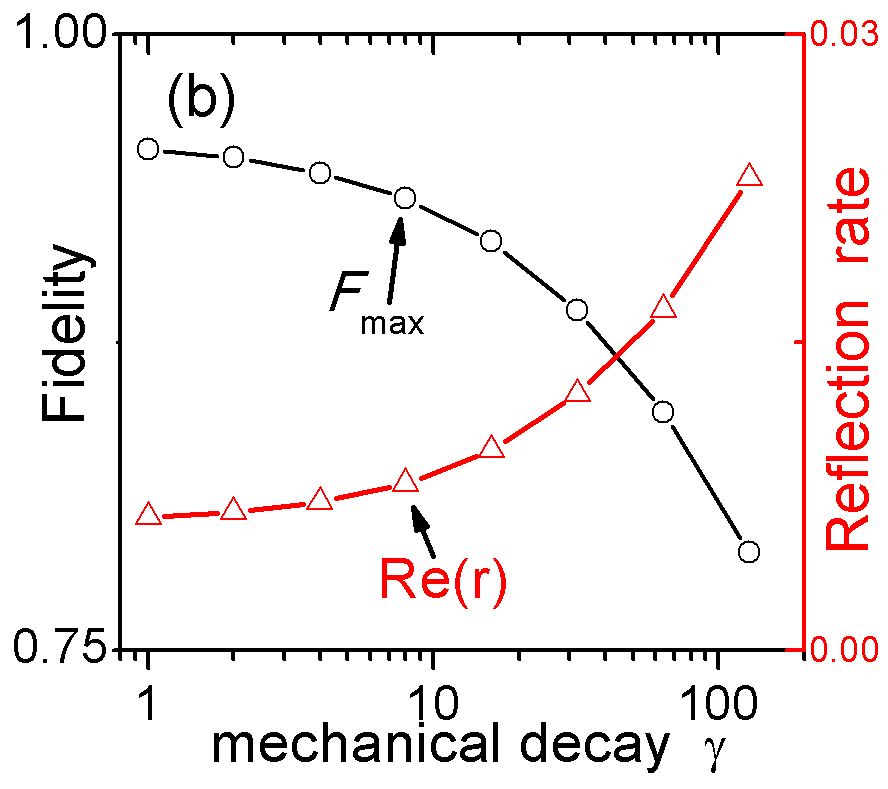}

\caption{(Color online) (a) The highest fidelity $F_{\rm{max}}$,
dispersion rate $\text{Im}(r)$, and reflection rate $\text{Re}(r)$ versus (a) the detuning
$\Delta_{d}$ and (b) the mechanical decay rate $\gamma$. In plot
(b), since $\text{Im}(r)\ll\text{Re}(r)$, $\text{Im}(r)$ is not
shown. Other parameters are the same as those in Fig.~\ref{fig3}.}
\label{fig5}
\end{figure}

\section{Detecting dark-state trapping}

Once the whole system is trapped into the dark superposition
states, the sideband and resonant transitions have equal
amplitudes but opposite signs, leading to destructive
interference, which is similar to the coherent population process
in $\Lambda$-type atomic systems~\cite{arimondo1996v}. After the
dark state is trapped, we can observe electromagnetically induced
transparency
(EIT)~\cite{Fleischhauer05,Anisimov2008,Anisimov11,Peng2014,Gu16}.
Without adding any other auxiliary detecting setups, one can
measure the reflected control fields to confirm the preparation of
the dark states~\cite{Murali04,Dutton06,Wang16e}. For the
resonant-frequency component $\omega_{2}$, the scattered current
amplitude can be expressed as $I_{\rm{sc}}=i\Gamma\langle\sigma
_{+}\rangle/\mu$, and the reflection coefficient is defined
as~\cite{Astafiev2010}:
\begin{equation}
  r(\omega_{2})=-\frac{I_{\rm{sc}}}{I_{2}}=\frac{i\Gamma\langle\sigma
_{+}\rangle}{(2\epsilon_{2})},
 \label{N8}
\end{equation}
for which real and imaginary parts are related to reflection and
dispersion, respectively. In experiments, the quadratic coupling
might be off-resonantly induced with detuning $\Delta_{d}$. In
Fig.~\ref{fig5}(a), we fix the resonant drive parameters, and show
how the highest fidelity $F_{\rm{max}}$ and the refection rate $r$
change with the sideband detuning $\Delta_{d}$. This is clearly
seen that, at the resonant sideband detuning $\Delta_{d}=0$, there
is almost no reflection for the resonant drive with
$\text{Re}(r)=\text{Im}(r)\simeq0$, and $F_{\rm{max}}$ reaches its
maximum. When $\Delta_{d}$ starts to bias from zero,
$F_{\rm{max}}$, starts to decrease, and both $\text{Im} (r)$ and
$|\text{Re} (r)|$ increase rapidly, indicating that the resonant
drive field is strongly reflected by the flux qubit. It can be
found that a Lorentzian dip occurs in $\text{Re}(r)$, while
$\text{Im}(r)$ follows a typical EIT dispersion curve around
$\Delta_{d}=0$.

In Fig.~\ref{fig5}(b), by considering the resonant case of
$\Delta_{d}=0$ [corresponding to the dip in Re($r$) in
Fig.~\ref{fig5}(a)], we plot $\text{Re}(r)$ and $F_{\text{max}}$
versus the CNT decay rate $\gamma$. [$\text{Im}(r)$ is not shown,
since $\text{Im}(r)\ll\text{Re}(r)$] It can be clearly found that,
when increasing $\gamma$, the highest fidelity $F_{\rm{max}}$
decreases rapidly, and the reflection coefficient $\text{Re}(r)$
also increases. Unfortunately, the reflection coefficient is not
sensitive as in the detuning case. Specifically, when
$\gamma=130~\rm{Hz}$, the highest fidelity is $F_{\rm{max}}\simeq0.80$,
while the reflection coefficient is only $\rm{Re}(r)\simeq0.024$, which
is too weak to be effectively measured in experiments. It is
because the dark state depends on the initial states and is not
unique, and the rapid mechanical decay results in the fidelity
decreasing quickly, while only a large $\gamma$ has a significant
effect on the dip of the reflection rate~\cite{Wang16e,Wang17}.
Thus, the error transitions caused by the thermal noise cannot be
observed with a high sensitivity from the optical response of the
qubit. However, if we can confirm that the mechanical decay is
extremely slow, observing the EIT of the control fields can also
be a strong indicator for the dark cat-state generation. A
detailed discussion of this method and its applications will be
presented elsewhere~\cite{Wang17}.

\section{Conclusions}

We proposed a novel hybrid quantum system, in which the mechanical
motion of a CNT strongly interacts with a flux qubit via a
longitudinal coupling. In such a system, the decoherence of the
qubit can be effectively suppressed. We showed how the ground and
squeezed ground states of the CNT can be achieved by sideband
cooling.

Moreover, by inducing a strong quadratic coupling in this hybrid
system, we can generate macroscopically-distinct superposition
states (Schr\"odinger cat-like states) of the mechanical mode by
taking advantage of the decay of the qubit. Since we consider dark
states and a high-quality-factor CNT, the superposition can live
long. We have shown that these cat states can be trapped in a dark state
assuming that the CNT dissipation is negligible
compared to the qubit dissipation. However, some experiments might
satisfy the opposite condition: the CNT dissipation being larger
than the qubit dissipation. Still, the original assumption could
be realized, e.g., by adding an extra noise to the qubit, while
keeping the CNT dissipation fixed. Finally, we showed how to
reveal the trapping of the dark superposition states by observing
the optical response of the control fields.

Our proposal can also be employed to demonstrate other
nonclassical mechanical effects, such as phonon
blockade~\cite{Liu10p,Adam16,Wang16} or generating macroscopically-distinct superpositions of more than two states (i.e., Schr\"odinger kitten states)~\cite{Adam90,Adam14}, and also might serve as a
nanomechanical quantum detector of weak forces or other weak
signals.

\begin{acknowledgments}
The authors acknowledge fruitful discussions with Dr. Dong Hou. XW
is supported by the China Scholarship Council (Grant No.
201506280142). AM and FN acknowledge the support of a grant from
the John Templeton Foundation. FN was partially supported by the
RIKEN iTHES Project, MURI Center for Dynamic Magneto-Optics via
the AFOSR Award No. FA9550-14-1-0040, the Japan Society for the
Promotion of Science (KAKENHI), the IMPACT program of JST,
JSPS-RFBR grant No 17-52-50023, and CREST grant No. JPMJCR1676.
\end{acknowledgments}
%

\end{document}